\documentclass[multphys,vecphys]{svmult}

\usepackage{makeidx}     % allows index generation
\usepackage{graphicx}    % standard LaTeX graphics tool
\usepackage{multicol}    % used for the two-column index
\usepackage{subfigure}
\def\WAT {\hbox{${\rm H}_2{\rm O}$}}   %H2O
\def\WATEIGH {\hbox{${\rm H}_2^{18}{\rm O}$}}   %H218O
\def\kms {\hbox{${\rm km\, s}^{-1}$}}

\def\percc {\hbox{{\rm cm}$^{-3}$}}    %cm-3
\def\cmsq  {\hbox{{\rm cm}$^{-2}$}}    %cm-2

\makeindex             % used for the subject index
\begin{document}

\title*{The line-of-sight distribution of water in SgrB2}
% Use \titlerunning{Short Title} for an abbreviated version of
% your contribution title if the original one is too long
\author{Claudia Comito \inst{1}
  \and 
  Peter Schilke\inst{1} 
  \and 
  Maryvonne Gerin\inst{2}
  \and
  Tom~G.~Phillips\inst{3}
  \and
  Jonas~Zmuidzinas\inst{3}
  \and
  Darek~C.~Lis\inst{3}}
\authorrunning{Comito et al.} 
\institute{Max-Planck-Institut f\"ur Radioastronomie, Auf dem H\"ugel 69,
     D-53121 Bonn, Germany
\texttt{ccomito@mpifr-bonn.mpg.de}
     \and
     Laboratoire de Radioastronomie Millim\'etrique, Observatoire de Paris and 
     Ecole Normale Sup\'erieure, 
     24 rue Lhomond, F-75231 Paris, CEDEX 05, France
     \and
     California Institute of Technology, Downs Laboratory of Physics 320-47, 
     Pasadena, CA 91125, USA
} 
%
% Use the package "url.sty" to avoid
% problems with special characters
% used in your e-mail or web address
%
\maketitle
\section{\WAT\ in SgrB2}\label{intro}
The SgrB2 cloud is one of the most massive star forming regions in our
Galaxy.  It has many unique characteristics, among them an exceptional
chemistry.  Several species (FeO, \cite{walmsley2002}; NH$_2$,
\cite{vandishoeck1993}; HF, \cite{neufeld1997}; C$_3$,
\cite{giesen2001}) have, in spite of searches elsewhere, been detected
only toward this source.  One possible explanation for this enigmatic
chemistry is the existence of a layer of hot gas, which is thought to
be produced by a shock.  Since this layer has the same velocity as the
ambient gas in SgrB2, it has been proven difficult to assess its
importance for the chemistry of many species, with a few exceptions.
In particular its importance for the water chemistry has been a matter
of recent debate (\cite{ceccarelli2002}; \cite{neufeld2003}).  While
its temperature and density are well known (cf. \cite{huette1995}),
its column density and width have remained elusive.  In this work
(illustrated in detail in \cite{comito2003}) we have been able to
determine its importance for water chemistry and give constraints for
column density and spatial width, by modeling the HDO and \WATEIGH\ 
emission and absorption.

Water is known to be a fundamental ingredient of the interstellar
medium.  An ubiquitous tracer of shock-heated gas, it dramatically
influences the chemistry in shocked regions (cf.  \cite{neumel87};
\cite{bergin1998}). In general, it acts as a a major coolant in
star-forming clouds (\cite{ceccarelli1996}).  Although direct
observation of non-masing water lines from ground-based observatories
is made extremely difficult by atmospheric absorption, \WAT\ 
abundances can be estimated via observations of isotopomers, such as
\WATEIGH\ (e.g.  \cite{phillips1978}; \cite{jacq1988} and
\cite{jacq1990}; \cite{gensh1996}), and, in recent years, by the
availibility of satellites such as ISO (e.g.  \cite{cernicharo1997})
and notably SWAS (cf. \cite{neufeld2000}), but also by airborne
observations of H$_2^{18}$O (e.g. \cite{zmuidzi95a}).  However, the
deuterated counterpart of water, HDO, presents many features
observable from ground in the cm, mm and submm wavelength atmospheric
windows (cf. \cite{henkel1987}; \cite{jacq1990} and \cite{jacq1999};
\cite{pardo2001}), and is also used as a tracer of water abundance
under the assumption that both the deuterated and non-deuterated
species be spatially coexistent, and that their abundance ratio remain
constant throughout the studied region.

\begin{figure}[htbp]
%  \centering 
  \subfigure{
    \includegraphics[angle=-90,bb=32 51 570 769,clip,width=6cm]{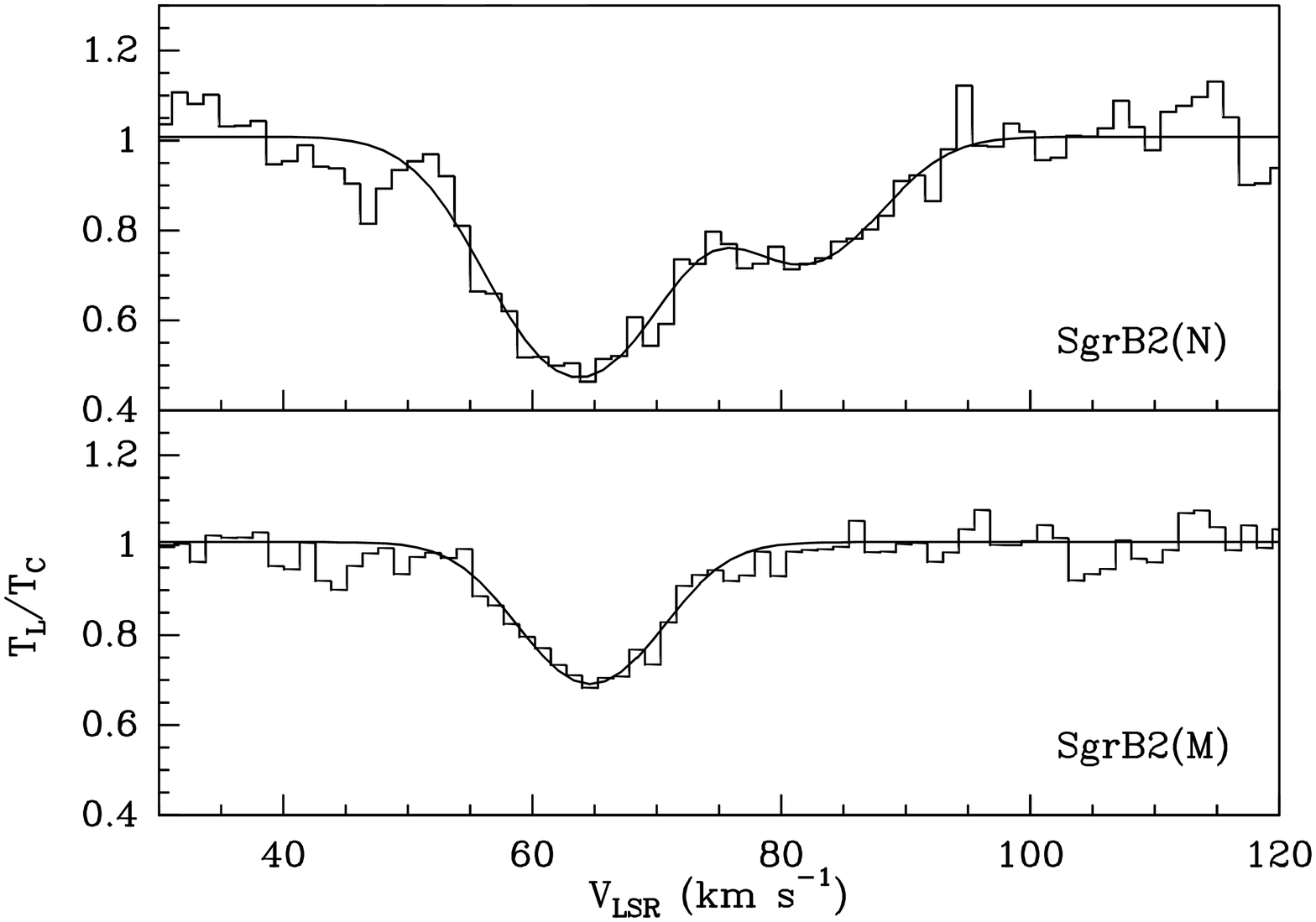}}
  \subfigure{
    \includegraphics[angle=-90,bb=32 51 570 769,clip,width=6cm]{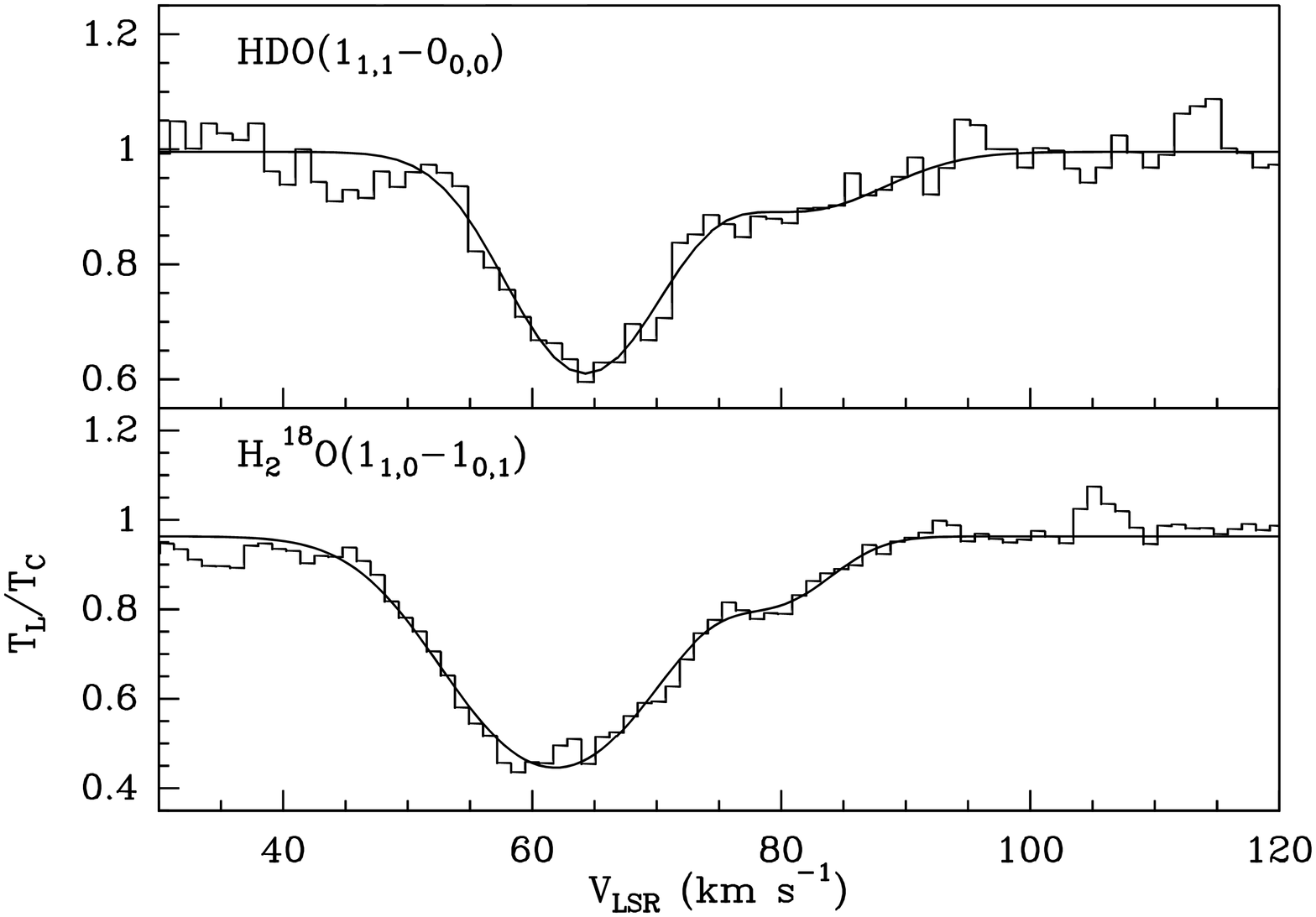}}
  \caption{{\it Left panel:} the 894-GHz HDO ($1_{1,1}-0_{0,0}$) transition, observed in 
    absorption against the continuum emission of SgrB2(M) and of
    SgrB2(N). {\it Right panel:} comparison between the 894-GHz
    HDO($1_{1,1}-0_{0,0}$) and the 547.7-GHz
    o-\WATEIGH($1_{1,0}-1_{0,1}$) absorption features observed towards
    SgrB2. 
%The HDO spectrum is the sum of the scans relative to the
%    894-GHz absorption towards SgrB2(M) and SgrB2(N) separately (see
%    left panel), observed with a $10\arcsec$ beam. The o-\WATEIGH\ 
%    spectrum shows the absorption observed by \cite{neufeld2000} with SWAS ($3\farcm3
%    \times 4\farcm5$ beam).  
The $y$-scale gives the line-to-continuum
    ($\rm{T_L/T_C}$) ratio.
%The absorption towards
%  SgrB2(M) shows one velocity component only, around 65 \kms, while
%  the gas in front of SgrB2(N) shows an additional component around 81
%  \kms. Such behaviour has been found for other molecular species as
%  well (e.g.  H$_2$CO, see Mart\'{\i}n-Pintado et al.
%  \cite{mpintado1990}).
}\label{fig:sgrb2m_n}
\end{figure}

We report the detection of the 894-GHz ground-state $1_{1,1}-0_{0,0}$
transition of HDO \emph{in absorption} against the background
continuum sources SgrB2(M) and SgrB2(N). The radial velocity of the
observed HDO features suggests that the absorption is directly
connected to the SgrB2 complex.  The fact that, i), we observed a
ground-state transition, and ii), we observed it in absorption,
allowed us to estimate, indipendently of assumption on the physical
state of the gas, the HDO column density, which we found to be of
1.2$\times$10$^{13}$ \cmsq\ towards SgrB2(M), and of
2.4$\times$10$^{13}$ and 1.2$\times$10$^{13}$ \cmsq\ towards the 65-
and the 81-\kms\ velocity components of SgrB2(N) respectively.
Moreover, using the H$_2^{18}$O SWAS data (\cite{neufeld2000}), we were able to
estimate a lower limit of 5$\times10^{-4}$ for the [HDO]/[H$_2$O]
ratio towards the SgrB2 complex.
  
The above mentioned results hold under the assumption that all the
absorbing water lies in the warm envelope. \cite{ceccarelli2002}
propose that H$_2$O might instead be located in the hot gas layer that
surrounds the whole SgrB2 complex, implying that gas-phase water in
the warm envelope does not contribute at all to the observed
absorption features. In fact, while this hypothesis is reasonable for
H$_2$O (the gas-phase production of water is triggered by high
temperatures), it would not be applicable to HDO (the same high
temperatures inhibit its enhancement). We therefore think it possible
that the gas observed in HDO and that seen in \WAT\ absorption be
located in distinct regions.

\section{Modeling the distribution of water}

SgrB2 is often adopted as a template region for the study of sites of
massive star formation. For this purpose, it is important to evaluate
to what extent the warm envelope and the hot layer \emph{separately}
contribute to the observed HDO and \WAT\ absorption. In particular, in
order to better understand the nature of the hot layer and possibly
the mechanism that created it, it is vital to learn more about its
chemical and physical characteristics, and especially about its water
content. A precise determination of the column density of water would
set a tight constraint on the physical models that attempt to identify
the heating mechanism responsible for the high gas temperatures.
Unfortunately, since both gas components show roughly the same radial
velocity, it is not possible to separate, by purely observational
means, the ground-state water absorption produced in the warm envelope
from that produced in the hot layer. However, a number of HDO and
\WATEIGH\ transitions have been observed towards the SgrB2 cores with
a variety of instruments.

Our 893-GHz HDO data, together with the published HDO and \WATEIGH\ 
observations carried out at mm and submm wavelengths, provide strong
observational constraints and enable us to model, in a self-consistent
manner, the HDO and \WAT\ abundance in all three components of the
SgrB2 cloud. In detail:

\begin{itemize}
\item[{\it a)}] the observed \emph{emission} lines of p-\WATEIGH\ at
  203 GHz (SgrB2(N), \cite{gensh1996}), and of HDO at 143, 226, 241
  GHz (SgrB2(M) and (N), \cite{jacq1990}) constrain the HDO and water
  abundance and the H$_2$ density in the hot-core-type components;
\item[{\it b)}] the 893-GHz HDO \emph{absorption} feature observed
  towards SgrB2(M) and N, together with the [HDO]/[\WAT] ratio
  estimated at point {\it a)}, help us set the HDO and \WAT\ abundance
  in the warm envelope;
\item [{\it c)}] finally, having estimated the column density of water
  in the warm envelope, it is possible to assess whether or not the
  quantity of non-deuterated water in the warm envelope can be
  responsible for the \WATEIGH\ absorption observed at 548~GHz
  (\cite{neufeld2000}, \cite{zmuidzi95a}).
\end{itemize}
We use the radiative transfer code described by \cite{zmuidzi95b},
with a few modifications (see \cite{comito2003} for details), to
reproduce the intensities observed for the features.  The model
consists of two spherically symmetric components representing the hot
core and the warm envelope (see \S~\ref{intro}).  Most of the spectral
line data are reproduced within errors of $\sim 30$\%, with two
exceptions:
\begin{itemize}
\item[1)] The 226-GHz HDO emission line observed towards SgrB2(M) is a
  factor of 2 weaker, and the 143-GHz HDO line in SgrB2(N) is 4 times
  stronger than predicted by the model. Such discrepancies are likely
  to be due to the intrinsic chemical differences between the two hot
  cores (cf.  \cite{miao1997} and references therein), which has not
  been taken into account in our model.
\item[2)] Most interestingly, the predicted ground-state o-\WATEIGH\ 
  absorption at 548 GHz is much shallower than observed
  (\cite{neufeld2000}, cf.  Fig.~\ref{fig:sgrb2m_n}): \emph{the
    abundance of non-deuterated water in the warm envelope is not
    sufficient to produce the absorption feature observed at 548 GHz.}
\end{itemize}

This is a very important result: our predictions confirm, at least
from a qualitative point of view, the hypothesis of
\cite{ceccarelli2002} that the most important contribution to the
observed \WATEIGH\ absorption come from the foreground hot gas layer.
Having predicted the contribution of the warm envelope to the
\WATEIGH\ absorption, it is possible to estimate the spatial width and
\WAT\ abundance necessary for the hot layer to produce the observed
feature, given the observational constraints mentioned in section
\S~\ref{intro}.  We use the radiative transport equation to calculate
the total intensity emerging from the hot layer, given the background
emission produced by hot cores and warm envelope.  The data are best
reproduced when the o-\WATEIGH\ column density in the hot layer is
$\sim 10^{14} \, \cmsq$. Assuming a [$^{16}$O]/[$^{18}$O] ratio of
261$\pm$20 (\cite{whiteoak1981}) and an ortho/para ratio of 3, we can
estimate the \WAT\ column density in the hot gas component to be of
order $\sim 3.5 \times 10^{16}$~\cmsq. If all the atomic oxygen not
locked in CO is bound in gas-phase water, then the \WAT\ abundance in
this region is about $5\times10^{-4}$~\percc, and the spatial width of
the hot layer is only 0.02~pc.
%
%
% BibTeX users please use
% \bibliographystyle{}
% \bibliography{}
%
% Non-BibTeX users please follow the syntax
% the syntax of "referenc.tex" for your own citations
%%%%%%%%%%%%%%%%%%%%%%%% referenc.tex %%%%%%%%%%%%%%%%%%%%%%%%%%%%%%
% sample references
% "physics"
%
% Use this file as a template for your own input.
%
%%%%%%%%%%%%%%%%%%%%%%%% Springer-Verlag %%%%%%%%%%%%%%%%%%%%%%%%%%

%
% BibTeX users please use
% \bibliographystyle{}
% \bibliography{}
%
% Non-BibTeX users please use

%%%%%%%%%%%%%%%%%%%%%%%%%%%%%%%%%%%%%%%%%%%%%%%%%%%%%%%%%%%%%%%%%%%%%%  }

%%%%%%%%%%%%%%%%%%%%%%%%%%%%%%%%%%%%%%%%%%%%%%%%%%%%%%%%%%%%%%%%%%%%%%

\printindex
\end{document}